\documentclass[letterpaper, 10 pt, conference]{ieeeconf}

\usepackage{amssymb}
\usepackage{amsmath}
\usepackage{romannum}
\usepackage{graphbox,graphicx}
\usepackage{float}
\usepackage{cite}
\usepackage{hyperref}
\usepackage{url}
\usepackage{calrsfs}
\usepackage{mathrsfs}
\usepackage{subfig}
\usepackage{xcolor}
\usepackage{enumerate}
\usepackage[font=footnotesize,labelsep=period]{caption}
\usepackage{siunitx}
\usepackage{algorithm}
\usepackage{algpseudocode}
\usepackage{cleveref}

\usepackage{titlesec}
\titlespacing*{\section} {0pt}{1ex}{0ex}

\title{\LARGE \bf Fault Detection and Tolerant Control for Aero2 2D0F Two-rotor Helicopter}

\author{Khalid Kabir Dandago,
	Long Zhang and Wei Pan 
}

\IEEEoverridecommandlockouts
\begin{document}

	\maketitle
	\thispagestyle{empty}
	\pagestyle{empty}
	\begin{abstract}
 Stability and satisfactory performance are key control requirements for any Unmanned Aerial Vehicle (UAV) application. Although conventional control systems for UAVs are designed to ensure flight stability and safe operation while achieving a desired task, a UAV may develop different types of flight faults that could lead to degradation in performance or, worse, instability. Unsatisfactory performance or instability of a UAV poses threats to lives, properties, and the flying environment. Therefore, it is crucial to design a system that can detect the occurrence of faults, identify the location at which the fault occurs, determine the severity of the fault, and subsequently use this information to accommodate the fault so that the vehicle can continue to operate satisfactorily. Even though performance analysis of faults is crucial in selecting the best strategies for fault detection and tolerance, little has been done in this regard, especially with real systems. Therefore, this paper analyzed the performance of a 2-degree-of-freedom (2DOF) bi-rotor helicopter’s control system in the presence of various actuator faults. Results from different faulty conditions indicate that faults degrade the performance of a conventional control system on UAVs and introduce vibrations into the system. These findings are more apparent when a fault leads to asymmetry or imbalance of the system. However, further experiments have shown that proper fault diagnosis and accommodation methods could help maintain satisfactory performance of the system in the presence of faults.\\ \\
 \textit{Keywords\textemdash Fault Accommodation, Fault Analysis, Fault Diagnosis, Fault Detection, Helicopter, Unmanned Aerial Vehicle.}
\end{abstract}	
 \section{Introduction}
	\label{sec1}
Unmanned Aerial Vehicles (UAVs), also referred to as drones, are becoming integral part of our daily activities. Their vast functionality in military operations, precision agriculture, traffic regulation, building, aerial photography, among other applications \cite{1tezza2019state,2outay2020applications,3sledz2021applications} have contributed to their rapid demand. Also, advancements in design techniques, as well as sensor and communication technologies have led to their impressive evolution \cite{4alladi2020applications}.

The three main types of UAVs based on aero-structural configuration are flapping-wing, fixed-wing and rotorcraft UAVs \cite{5abdelmaksoud2020control}. Comparatively, fixed-wing UAVs have higher endurance, range coverage and cruise speed \cite{6dundar2020design}. However, rotorcraft, which is propelled by a number of motors (depending on its configuration), offers affordable performance-to-price ratio, simplicity, low maintenance cost, and high agility \cite{7liu2019collision,8narvaez2020autonomous}. Due to the various benefits of helicopters and quadcopters, they are the most researched rotorcrafts in the literature \cite{5abdelmaksoud2020control}. While a helicopter is configured to operate with two rotors, quadcopter is equipped with four motors as its actuators.

Helicopters and quadcopters, as well as other UAVs, require robust control strategies to efficiently handle flight-related scenarios such as obstacle avoidance, trajectory tracking, internal/external disturbances, and faulty conditions. 
\cite{5abdelmaksoud2020control} classified control strategies into four categories, which are: 1) Linear 2) Nonlinear 3) Artificial intelligent and 4) Hybrid control schemes. Each of these schemes has its own pros and cons, and selection of any of them usually depends on functional requirements and trade-offs.

\begin{figure}[tp]
	\centering
 \includegraphics[width=9cm,height=20cm,keepaspectratio]{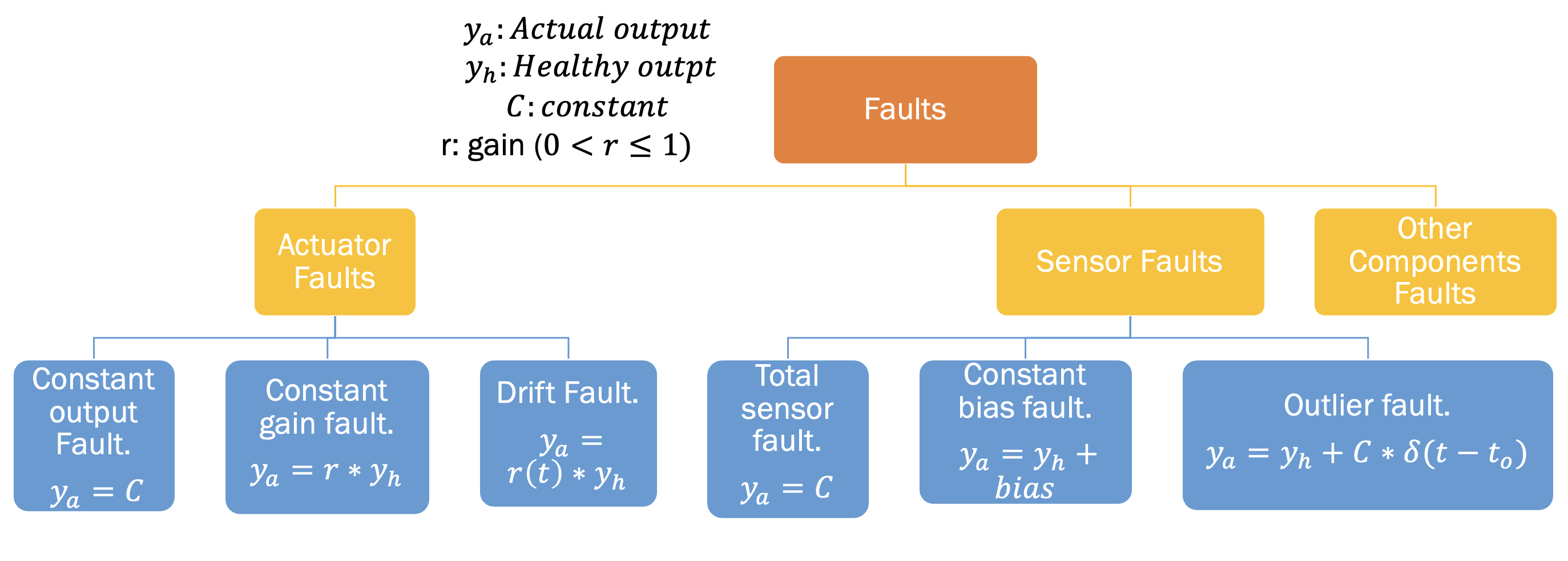}
	\caption{Classification of different types faults that may affect the operation of a UAV \cite{22qi2013literature}. }
	\label{image1}
\end{figure}

A UAV may develop different types of fault during its flight operations. UAV faults are majorly categorized into three: actuator faults, sensor faults and other components faults \cite{22qi2013literature}. The paper further defines actuator fault as the partial or total loss of an actuator’s control action. Moreover, actuator fault was subdivided into constant output, constant gain, and drift faults. Among the three, constant gain fault is more prevalent. On the other hand, sensor fault was classified into total sensor fault, constant bias fault, drift fault and outlier fault. These divisions are further illustrated by \cref{image1}.

\begin{figure}[tp]	\centering
 \includegraphics[width=9cm,height=20cm,keepaspectratio]{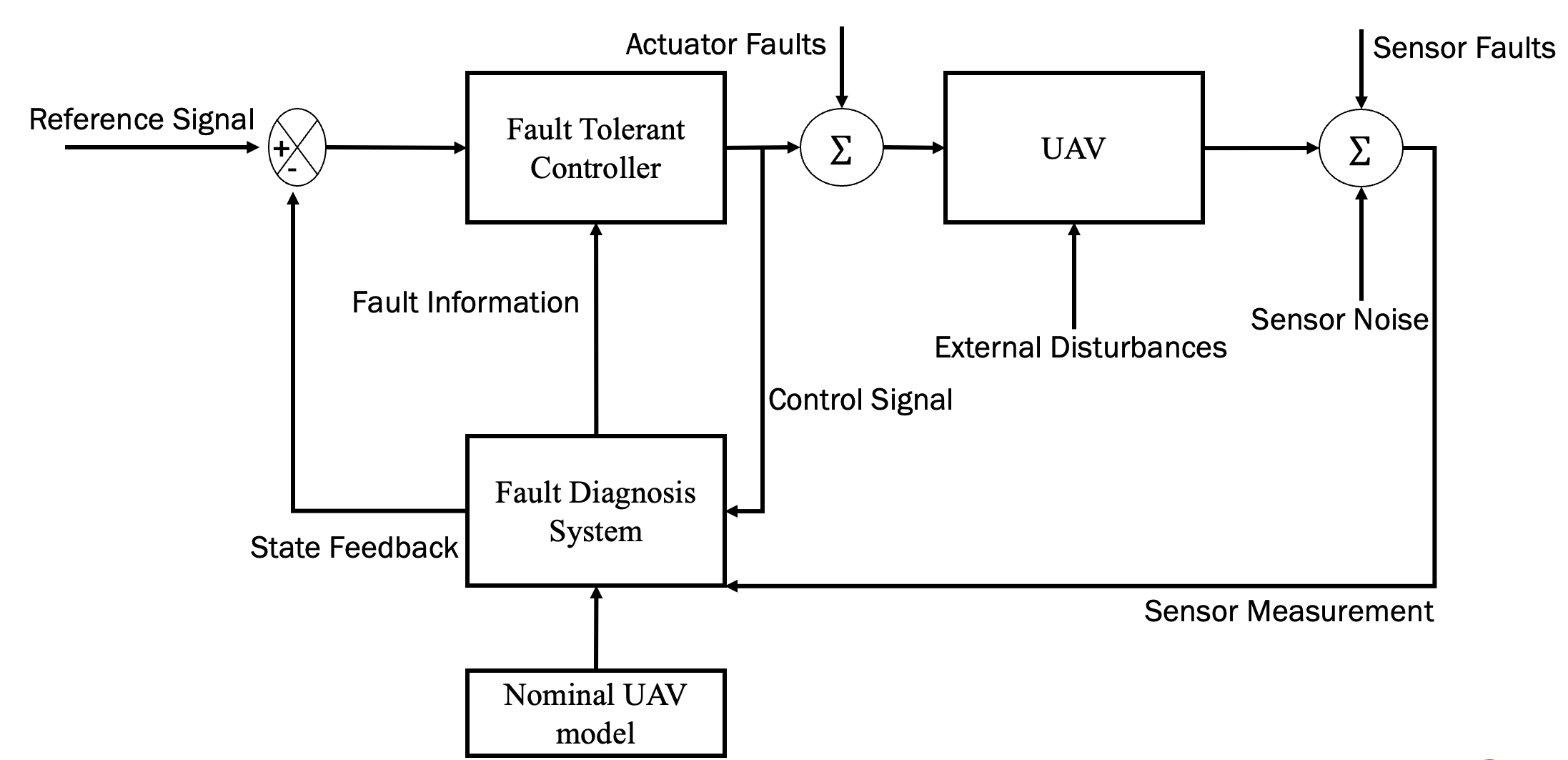}
	\caption{Block diagram that illustrates the design of a fault diagnosis and tolerance control system.}
	\label{image2}
\end{figure}

In the advent of any of these faults, conventional feedback control systems might not be able to maintain the stability and satisfactory performance of the vehicle. Hence, the occurrence of fault in such settings might be catastrophic.  Therefore, to ensure flight safety whenever faults occur, it is essential to design a flight control system that is robust enough to handle the occurrence of a fault. Such control system should consist of diagnosis and tolerance subsystems. The diagnosis subsystem is responsible for determining the occurrence of fault, knowing the location at which the faults occurs, and determining the severity of the fault.  Subsequently, the tolerance subsystem would use the information gathered by the diagnosis subsystem to reconfigure the employed controller and accommodate the fault. \cref{image2} further illustrates the synergy between fault diagnosis and fault tolerance subsystems in handling faults on UAVs.

Fault tolerant control systems for UAVs are widely classified into two: active and passive fault tolerant control systems (AFTCS and PFTCS) \cite{23wang2020dual}. PFTCS are designed to be robust against presumed faults. Therefore, they do not need a Fault Diagnosis (FD) system to detect, isolate and estimate the magnitude of a fault \cite{24wang2020active}. Also, they don’t require the reconfiguration of the control scheme adopted to handle an existing fault. As such, they have limited fault handling capabilities as they can only respond to faults that were considered during design. Conversely, AFTCS are designed to actively handle faults as they arise during flight operation. This type of configuration requires FD scheme that will detect the occurrence of fault, the location at which it occurs, and estimate fault parameters and/or post-fault models \cite{25baldini2020actuator}. The obtained information is then used to reconfigure the controller to achieve overall system stability and satisfactory performance \cite{26zhao2021data}.

The relatively simple structure and less computational burden of PFTCS have garnered it noticeable interest in the literature. The strong capability of handling uncertainties and disturbances, as well as being insensitive to parameter uncertainties have made Sliding Mode Control (SMC) an excellent choice for PFTCS \cite{23wang2020dual,19gao2022adaptive}. The performance of a sensor-based incremental and model-based dynamic inversion control schemes that were based on sliding mode disturbance observers was compared on fault tolerant control of a quadrotor \cite{27wang2019quadrotor}. The sensor-based method was found to outperform its model-based counterpart due to its lower uncertainty and less dependence on the vehicle’s model.

Another controller that was successfully used to implement PFTCS for UAVs is the backstepping controller \cite{21zhang2012fault,28avram2017nonlinear}. Furthermore, A passive fault tolerant control scheme was developed using control allocation scheme to satisfactorily handle the complete failure of one rotor during the flight operation of a quadrotor \cite{29merheb2017emergency}. Upon occurrence of such fault, the faulty rotor is exempted from control; therefore, the vehicle operated like a tri-copter and PID control signals were adequately distributed to the three healthy rotors. Although the stability of the vehicle was maintained, degradation in performance was obvious due to the non-symmetrical nature of a tri-copter, activation delay and loss of yaw control.

A key component of the AFTCS is the FD scheme which generates the fault information that is utilized by the adopted control algorithm to adequately reconfigure itself to maintain a system’s stability and an acceptable performance in all faulty conditions. FD methods are generally categorized into three: 1) hardware redundancy method 2) analytical method 3) signal processing method \cite{30fourlas2021survey}. Multiple identical components are used in the hardware redundancy technique, for the same function, to easily detect and isolate any component that develops a fault \cite{31guo2018fault}. Although a highly reliable method, it is not practical for rotary UAVs due to the increase in power demand \cite{32kirschstein2020comparison} and the decrease in agility \cite{33diehl2017conceptual} that it would introduce to the system. The analytical method makes use of a system’s model/information to diagnose a fault \cite{34lyu2019analytical}. This method is further classified into model-based and knowledge-based analytical methods \cite{30fourlas2021survey}. In the model-based method, the residual signal between a sensor measured signal and the corresponding signal estimated by a system’s model is used to extract fault information from a system \cite{38nian2020robust}. On the other hand, knowledge-based method depends on past system performance data and expert knowledge to execute diagnostic procedure \cite{35zhang2021fault}. Finally, the signal processing method depends on measured signals, specialized processing tools and established knowledge on the symptoms of an unhealthy/healthy system to determine the faulty conditions of the system \cite{30fourlas2021survey,31guo2018fault,45avram2017quadrotor}.

Recurrent neural network \cite{24wang2020active}, Gain-scheduling PID controller \cite{21zhang2012fault}, Model predictive controller \cite{37ali2020fault}, Dynamic output feedback control algorithm \cite{38nian2020robust}, and nonlinear dynamic inversion method \cite{39sun2020incremental} are some of the control methods that were used for the design of AFTCS. Furthermore, feedforward neural network and a nonlinear observer were used to protect a UAV against false data injection through cyber-attacks \cite{42abbaspour2017adaptive}.

While numerous fault diagnosis and tolerance control systems have been developed and tested on UAVs, typically through simulations, only a few studies have analyzed the performance of faulty aircraft systems operating with conventional controllers. Such analyses are paramount in determining which diagnosis and tolerance methods to utilize for a particular system. Furthermore, analysis and diagnosis on real systems have been limited to a few faulty conditions. Therefore, this study analyzed the performance of various faulty conditions on the propeller blades of a 2-degree-of-freedom (2DOF) helicopter. Additionally, a fault diagnosis and accommodation scheme for the analyzed fault were proposed. Succeeding contents of the paper are presented as follows: \cref{sec2} explains the experimental set-up and the material/methods that were used to conduct the research work. \cref{sec3} presents results obtained from the experiments. \cref{sec4} discusses the results, and finally, concluding remarks were given in \cref{sec5}.

	\section{Technical background}
	\label{sec2}
\subsection{Robotic Platform}
\begin{figure}[tp]	\centering
 \includegraphics[width=9cm,height=20cm,keepaspectratio]{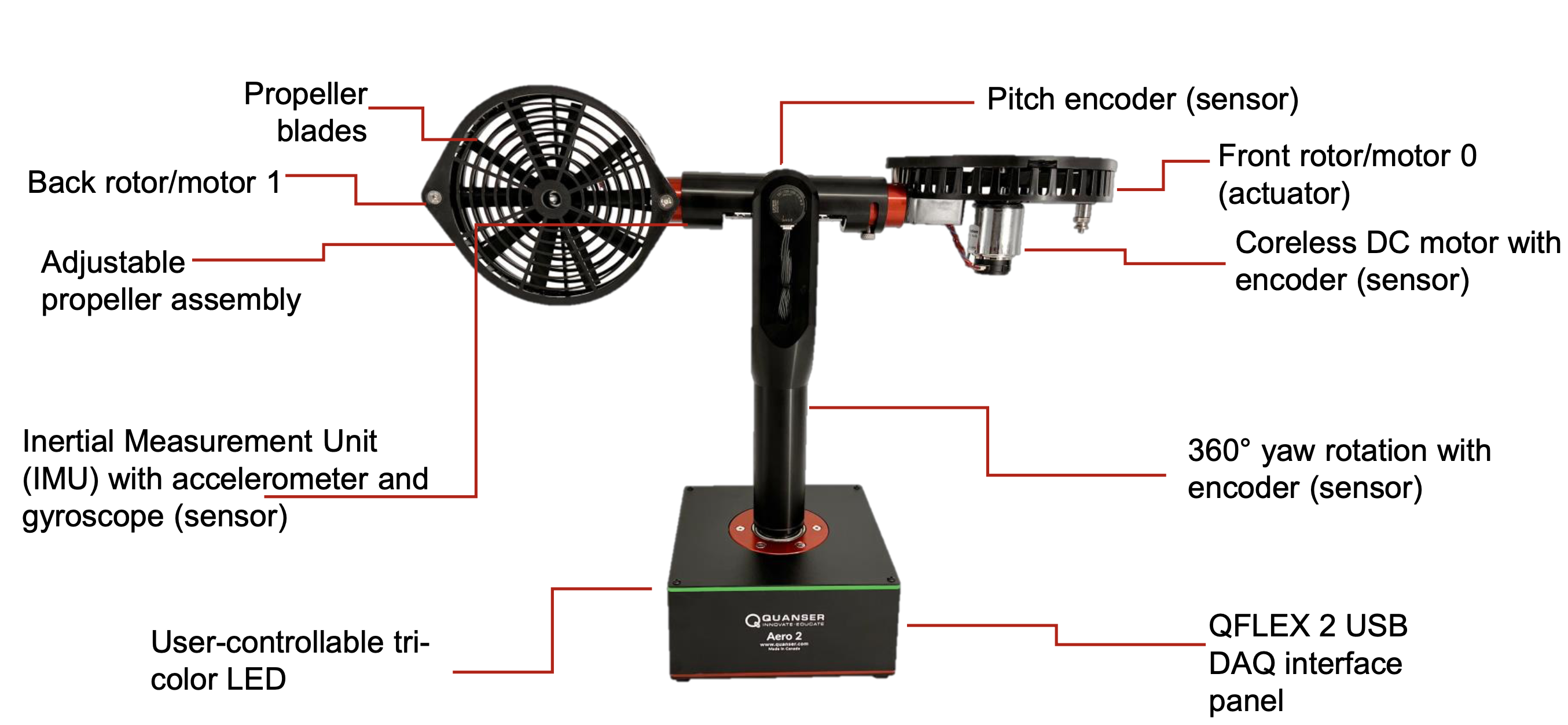}
	\caption{2-DOF helicopter configuration of the Aero 2 equipment [image by
Quanser Inc.]}
	\label{image3}
\end{figure}

Quanser's Aero2 aerospace system was used for the research work. Aero 2 is a reconfigurable dual-rotor aerospace system that could be used to emulate the operation of a 1-DOF VTOL, 2-DOF helicopter and a half quadrotor. The ability to lock/unlock its two axes—Pitch and yaw axes—and configure its thrusters vertically or horizontally gives users the flexibility of mimicking different aerospace systems. The 2-DOF helicopter setting of the system, which is shown in \cref{image3}, was used for conducting experiments. 

\subsection{Modelling of Aero2 2-DOF helicopter dynamics}
Although the pitch and yaw motions are primarily influenced by motor0 and motor1 of the system, respectively, coupling exists between the actions of the two motors due to reactive torques. The simplified Newtonian linear equation of motions of the 2 DOF helicopter are given by \cref{eq1} and \cref{eq2}. The model assumes that the thrusts generated by the motors are directly proportional to applied voltages. Even though the model takes gyroscopic reactive torques into account, it does not capture all system dynamics such as nonlinearities. Therefore, it is meant for designing linear control systems.

\begin{equation} \label{eq1}
\tau_{p}= K_{pp}V_{p}D_{t}+K_{py}V_{y}D_{t}=J_{p}\ddot\theta + D_{p}\dot\theta + K_{sp}\theta 
\end{equation}
\begin{equation} \label{eq2}
\tau_{y}= K_{yy}V_{y}D_{t}+K_{yp}V_{p}D_{t}=J_{p}\ddot\psi + D_{y}\dot\psi
\end{equation}

Some of the parameters in \cref{eq1} and \cref{eq2} are given in the system’s specification, while others were experimentally obtained. The values of the parameters are given by \cref{table1}.

\begin{table}[tp]
\centering
\caption{Model parameters}
\begin{tabular}{|c|c|}
\hline
\textbf{Parameters}& \textbf{Values} \\
\hline
$D_t (m)$  &	0.1674 \\
$K_{pp} (N/V)$ &	0.00321 \\
$K_{sp} (-)$ &	0.00744 \\
$D_p (\-)$	& 0.00199  \\
$J_p (Kg.m^{2}$	& 0.0232 \\
$K_{yy} (N/V)$ &	0.00610\\
$D_y (-)$ &	0.00192 \\
$J_y (kg.m^{2})$ &	0.0238  \\
$K_{yp} (N/V)$ &	-0.00319\\
$K_{py} (N/V)$ &	0.00137\\
\hline
\end{tabular}
\label{table1}
\end{table}
The terms in \cref{eq1} and \cref{eq2} are defined as:\\
$\tau_{p}$: Torque that generates the pitch motion\\
$K_{pp}$: Thrust-Force gain acting on the pitch axis from the main rotor\\
$D_{t}$: Distance between Aero 2 pivot and center of the rotors\\
$V_{p}$: Input voltage of the main motor\\
$J_{p}$: Total moment of inertia about the pitch axis\\
$D_{p}$: Damping about the pitch axis\\
$K_{sp}$: Stiffness about the pitch axis\\
$\tau_{y}$: Torque that generates the yaw motion\\
$K_{yy}$: Thrust-Force gain acting on the yaw axis from the rear rotor\\
$V_{y}$: Input voltage of the rear motor\\
$J_{y}$: Total moment of inertia about the yaw axis\\
$D_{y}$: Damping about the yaw axis\\
$K_{yp}$: Thrust-Force gain acting on the yaw axis from the front rotor\\
$K_{py}$: Thrust-Force gain acting on the pitch axis from the rear rotor

The developed state space model of the system from \cref{eq1}, \cref{eq2}, and \cref{table1} is given by \cref{eq3} and \cref{eq4}.

\begin{equation} \label{eq3}
\dot{X}= AX + BU
\end{equation}
\begin{equation} \label{eq4}
Y= CX
\end{equation}

where, \begin{equation*}A=\left[\begin{array}{cccc}
	 0 & 0 & 1 &0\\
      0 & 0 & 0 & 1 \\
      -0.3190 & 0 & -0.1164 & 0 \\
      0 & 0 & 0 &-0.1386\end{array}\right] \end{equation*}

     \begin{equation*} B= \left[\begin{array}{cc}
	 0 & 0\\
      0 & 0 \\
      0.0216 & 0.01154 \\
      -0.01336 & 0.052
\end{array}\right] \end{equation*}
\begin{equation*} C=\left[\begin{array}{cccc}
	 1 & 0 & 0 &0\\
      0 & 1 & 0 & 0 \\
      0 & 0 & 1 & 0 \\
      0 & 0 & 0 &1\end{array}\right] \end{equation*}
       \begin{equation*} X=[\theta, \psi, \dot\theta, \dot\psi]^{'} \end{equation*}
      \begin{equation*} U=[V_{p} \quad V_{y}]^{'} \end{equation*}
\subsection{Control Method}
\begin{figure}[tp]	\centering
 \includegraphics[width=9cm,height=20cm,keepaspectratio]{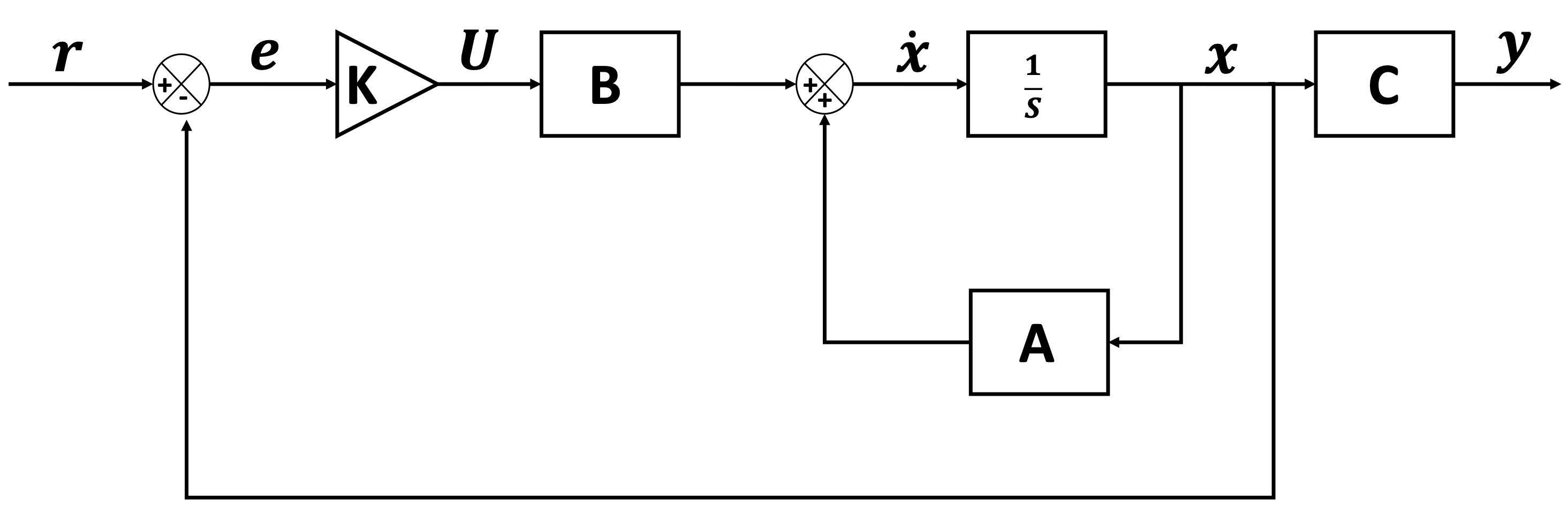}
	\caption{LQR control architecture}
	\label{image4}
\end{figure}

Linear Quadratic Regulator (LQR) was used for the pitch and yaw motions control of the 2DOF helicopter. LQR control is a multi-input multi-output (MIMO) full state feedback control system which computes an optimal control gain, K, of dimension $m*n$---where n is the number of states and m is the number of control outputs---by minimizing a cost function, J. The cost function constitutes of the tracking error of the system, as well as the amount of control input required to achieve the desired control objective. The formulation of the cost function is given by \cref{eq5}.
\begin{equation} \label{eq5}
\begin{split}
 &\min_{U=\mathbb{R}^m } \ J=\int_0^\infty e^TQe+U^TRU\\
 & subject \ to:\ \dot{x}=Ax+BU
 \end{split}
\end{equation}
From \cref{eq5}, e is the tracking error, U is the control input, while Q and R are the weighting matrices that are tuned to actualize the minimization preference of the tracking error and the control input, respectively. \cref{image4} visualizes the control structure of LQR.

It could be deduced from \cref{image4} that $U=K(r-x)$, where $K$ is obtained by solving \cref{eq3.22}.
\begin{equation} \label{eq3.22}
K=R^{-1}B^TP
\end{equation}
$P$ in \cref{eq3.22} is obtained by solving the Riccati equation, which is given by \cref{eq3.23}.
\begin{equation} \label{eq3.23}
A^TP+PA-PBR^{-1}B^TP+Q=0
\end{equation}
The weighting matrices that were used for this work are:
\begin{equation*}
Q=\left[\begin{array}{cccc}
	 150 & 0 & 0 &0\\
      0 & 75 & 0 & 0 \\
      0 & 0 & 0 & 0 \\
      0 & 0 & 0 &0\end{array}\right] \quad and \quad R=\left[\begin{array}{cc}
	 0.01 & 0\\
      0 & 0.01 
\end{array}\right]
\end{equation*}
\subsection{Faulty Conditions}
\begin{figure}[tp]	\centering
 \includegraphics[width=9cm,height=20cm,keepaspectratio]{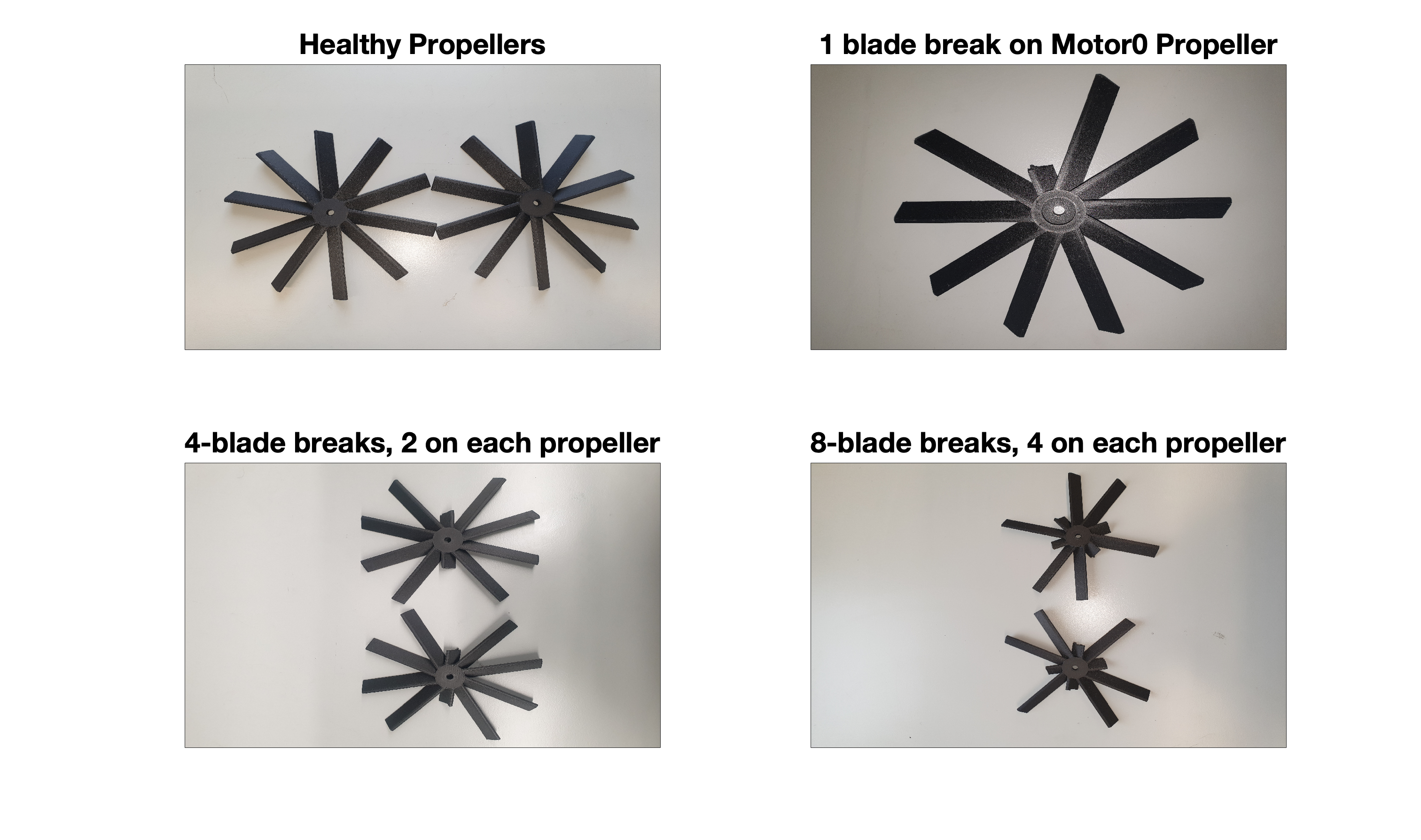}
	\caption{Blades breaks considered as faults for fault experiments.}
	\label{image5}
\end{figure}

The Aero2 2DOF helicopter operates with a pair of 8-blade propellers, with each mounted on one of its motors. Various types of breaks on the propeller blades were considered as faulty conditions for experiments. The system was operated with healthy propellers, one blade break, 2-blade breaks, 4-blade breaks, and 8-blade breaks. The breaks are visualized by \cref{image5}.
\subsection{Performance Metrics}
The impact of the faults on trajectory tracking of the system's pitch and yaw motions was assessed by measuring time response metrics such as rise time, overshoot, and steady-state error (SSE). Additionally, to examine the level of vibration introduced into the system by the faults, the standard deviation (SD) of the computed control input, as well as the SD of the pitch and yaw angles at steady state, were recorded. Furthermore, the open-loop natural frequency of the system was measured under the various fault conditions.
\subsection{Fault Diagnosis and Tolerance}
Breaks on propeller blades is an actuation problem that may lead to loss of control effectiveness. In several research works \cite{21zhang2012fault,28avram2017nonlinear,44zhou2021quadrotor}, such category of actuator faults has been modelled by \cref{eq4.11}:
\begin{equation} \label{eq4.11}
u_f^i=(1-\gamma_i)u_i
\end{equation}
Where $\gamma_i$ denotes the loss of control effectiveness in the $i_{th}$ actuator. $u_i$ is the control input generated by controller, and $u_f^i$ is the control input that the $i_{th}$ actuator operates on due to fault. 

$\gamma_i \in [0,1]$ is the fault parameter that defines the degree of healthiness of $i_{th}$ actuator; in the case of Aero2, we have motor0 and motor1. $\gamma_i=0$ signifies that an actuator is fully healthy, while $\gamma_i=1$ indicates complete failure of an actuator. $\Gamma$ is the defined as fault parameter vector and, for Aero2, it is given by $\Gamma=[\gamma_1 \quad \gamma_2]^{'}$.

As an unknown input to the system $\Gamma$ can be regarded as an augmented state to the conventional state-space model given by \cref{eq3} and \cref{eq4}. This generates an augmented state space model of the system that is given by \cref{eq9}.

The control input was modelled by \cref{eq4.11}. Therefore, the augmented system model becomes:
\begin{equation} \label{eq9}
\begin{split}
& \dot{X}=AX+B(1-\Gamma)U \\
&  Y=CX\\
&\Gamma=diag(\gamma_i)
\end{split}
\end{equation}
Hence,
\begin{equation} \label{eq11}
\begin{split}
& \dot{X}=AX+BU+N\Gamma \\
&  Y=CX\\
& N=-Bu, \quad u=diag(U) 
\end{split}
\end{equation}
The designed fault diagnosis scheme is a modification of the discrete state Kalman filter proposed by \cite{46zhong2018robust}. Therefore, \cref{eq11} was discretized to obtain \cref{eq12}.
\begin{equation} \label{eq12}
\begin{split}
& x_{k+1}=A_kx_k+B_ku_k+N_k\Gamma_k+\omega_k \\
&  y_k=C_kx_k+\nu_k
\end{split}
\end{equation}
Where $\omega_k$ and $\nu_k$ are uncorrelated zero-mean white gaussian noise sequences with covariances $Q_k$ and $R_k$, respectively. Considering $\Gamma_k$ as an augmented states vector, \cref{eq12} becomes \cref{eq13}.
\begin{equation} \label{eq13}
\begin{split}
& x_{k+1}^a=A_k^ax_k^a+B_k^au_k+\omega_k^a \\
&  y_k=C_k^ax_k^a+\nu_k^a
\end{split}
\end{equation}
where,
\begin{equation*}
A_k^a=\left[\begin{array}{cc} 
A_k &N_k\\ 
0 & I \end{array}\right], B_k^a=\left[\begin{array}{c} 
B_k\\ 
0  \end{array}\right], C_k^a=\left[\begin{array}{cc} 
C_k &0 \end{array}\right]
\end{equation*}
The filter has two steps: prediction and correction steps. 
\subsubsection{Prediction Step}
The mathematical formulation of the prediction step, which is also known as the control update step is given by \cref{eq14}-\cref{eq16}.
\begin{equation} \label{eq14}
\hat{x}_{k|K-1}^a=A_{k-1}x_{k-1}+B_{k-1}^au_{k-1}
\end{equation}
\begin{equation} \label{eq15}
P_{k|K-1}^a=A_{k-1}^aP_{k-1}^aA_{k-1}^{aT}+Q_{k-1}^a
\end{equation}
\begin{equation} \label{eq16}
K_k^a=P_k^aC_k^{aT}[C_k^aP_k^aC_k^{aT}+R_k^a]^{-1}
\end{equation}
Where K is the Kalman filter gain and P is the uncertainty matrix of the filter’s estimates.
\subsubsection{Correction step}
The estimate of the states and the uncertainty of the estimate is further corrected using the kalman filter gain using \cref{eq17} and \cref{eq18}.
\begin{equation} \label{eq17}
\hat{x}_{k|K}^a=\hat{x}_{k|K-1}^a+K_K^a(y_k-C_K^a\hat{x}_{k|K-1}^a)
\end{equation}
\begin{equation} \label{eq18}
P_{k|K}^a=P_{k|K-1}^a-K_K^aC_K^aP_{k|K-1}^a
\end{equation}
\subsubsection{Fault accommodation}
Upon estimation of the faults, the control input is adjusted using \cref{eq19}.
\begin{equation} \label{eq19}
u_f^i=(1-\gamma_i)u_i+\gamma_{estimated_i}u_i
\end{equation}
Where $\gamma_{estimated_i}$ is the fault parameter estimated for the $i^{th}$ rotor.

	\section{Results}
 \label{sec3}
 \subsection{Trajectory Tracking}
	
\begin{figure}[tp]	\centering
 \includegraphics[width=9cm,height=30cm,keepaspectratio]{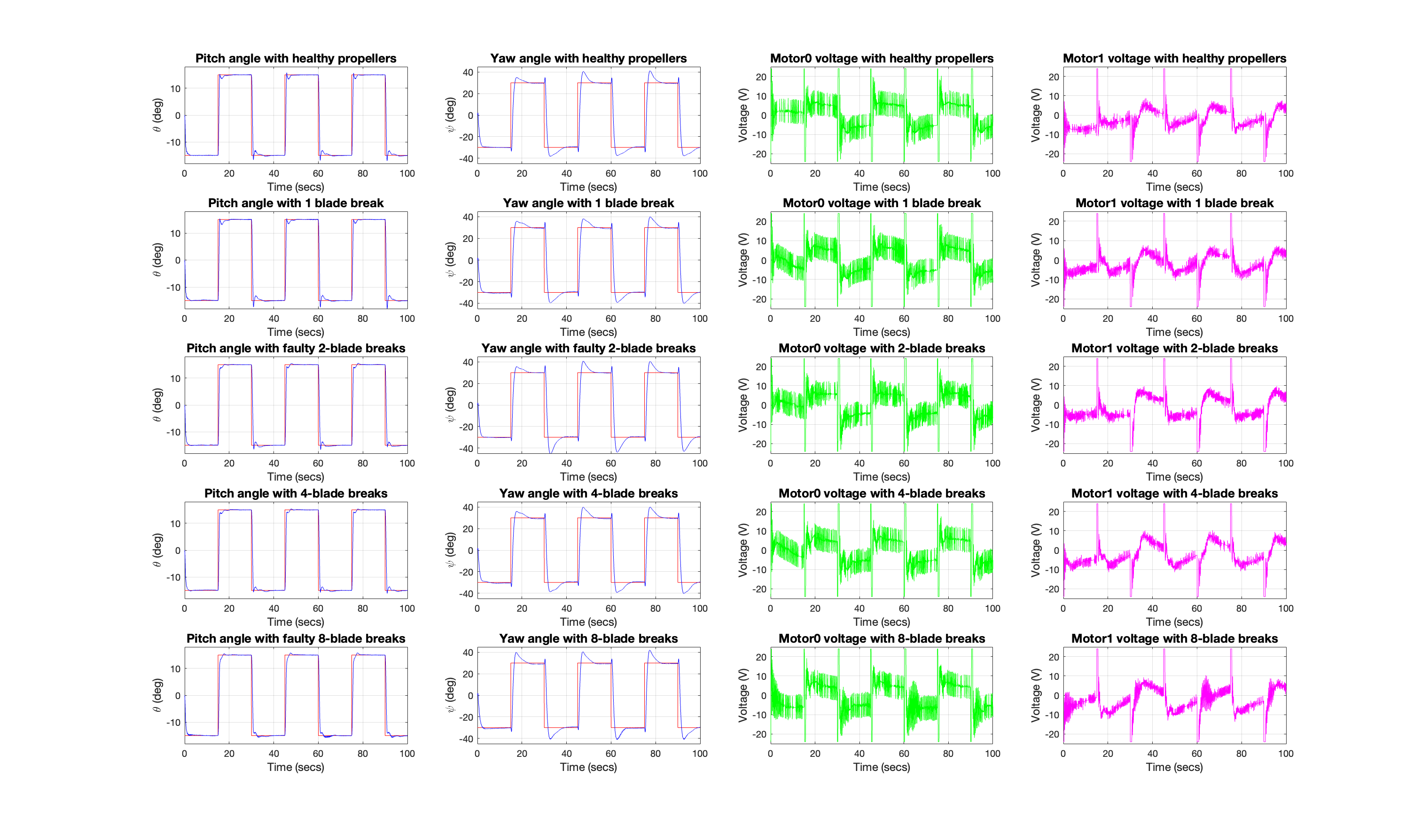}
	\caption{LQR Trajectory Tracking control of Aero2 2DOF helicopter's pitch and yaw motion under different fault conditions. Note: The 1 blade break is on Motor0, while half of the other breaks is on each Motor.}
	\label{image6}
\end{figure}

 The results of the trajectory tracking of the pitch and yaw motion of the real aerospace system, under different faulty conditions, is presented by \cref{image6}.
 The computed control inputs for the actuation of both Motor0 and Motor1 of the system is also presented in the figure. Although the controller has some level of inherent fault tolerance capability, upon the occurrence of fault, degradation in performance was observed in some of the metrics utilized for analysis. \cref{table2} and \cref{table3} presents the time response performance of the pitch and yaw tracking, respectively.

 \begin{table}[tp]
\centering
\caption{Pitch trajectory tracking performance for different fault conditions}
\begin{tabular}{|c|c|c|c|}
\hline
\textbf{Fault Conditions} & \textbf{SSE (\%)}& \textbf{Rise time (s)} & \textbf{Overshoot (\%)} \\
\hline
\text{Healthy Propellers}  &	0.04&0.64&5.29 \\
\text{1 blade break} &0.00&0.67&5.70 \\
\text{2-blade breaks} &0.09&0.76&0.83 \\
\text{4-blade breaks}	& 0.10&0.74&0.83  \\
\text{8-blade breaks}	& 0.09&1.18&2.25 \\

\hline
\end{tabular}
\label{table2}
\end{table}

\begin{table}[tp]
\centering
\caption{Yaw trajectory tracking performance for different fault conditions}
\begin{tabular}{|c|c|c|c|}
\hline
\textbf{Fault Conditions} & \textbf{SSE (\%)}& \textbf{Rise time (s)} & \textbf{Overshoot (\%)} \\
\hline
\text{Healthy Propellers}  &1.6&1.6&16.60 \\
\text{1 blade break} &2.22 &1.58&19.53 \\
\text{2-blade breaks} &0.68&1.59&18.65 \\
\text{4-blade breaks}	& 2.16&1.56&20.11  \\
\text{8-blade breaks}	& 2.5&1.44&32.72 \\

\hline
\end{tabular}
\label{table3}
\end{table}

To ascertain the level of vibration that was induced into the system as a result of the faults, the SD of the pitch and yaw angles' steady state values, as well as the SD of the steady state values of motor0 and motor1 voltages was observed. These values are presented in \cref{table4} and \cref{table5}.	

\begin{table}[tp]
\centering
\caption{Standard deviations of Pitch angle and Motor0 steady state values }
\begin{tabular}{|c|c|c|}
\hline
\textbf{Fault Conditions} & \textbf{Pitch angle SD (deg)}& \textbf{Motor0 Voltage SD (V)} \\
\hline
\text{Healthy Propellers}  &0.049&2.247 \\
\text{1 blade break} &0.07 &3.096 \\
\text{2-blade breaks} &0.049&2.014 \\
\text{4-blade breaks}	& 0.042&2.139  \\
\text{8-blade breaks}	& 0.044&2.187 \\

\hline
\end{tabular}
\label{table4}
\end{table}

\begin{table}[tp]
\centering
\caption{Standard deviations of Yaw angle and Motor1 steady state values }
\begin{tabular}{|c|c|c|}
\hline
\textbf{Fault Conditions} & \textbf{Yaw angle SD (deg)}& \textbf{Motor1 Voltage SD (V)} \\
\hline
\text{Healthy Propellers}  &0.156&1.413 \\
\text{1 blade break} &0.211 &1.962 \\
\text{2-blade breaks} &0.185&1.220 \\
\text{4-blade breaks}	& 0.148&1.380  \\
\text{8-blade breaks}	& 0.161&1.347 \\

\hline
\end{tabular}
\label{table5}
\end{table}
Furthermore, experiments were performed to notice the change in the natural frequency of the system's open loop response with different fault conditions. The result of the experiments is presented in \cref{table6}

\begin{table}[tp]
\centering
\caption{Natural frequency, $\omega_{n}$, of the system's Pitch angle with different fault conditions.}
\begin{tabular}{|c|c|}
\hline
\textbf{Fault Conditions} & \textbf{$\omega_{n}$ (rad/s)} \\
\hline
\text{Healthy Propeller}  &0.70 \\
\text{1 blade break} &0.73 \\
\text{2-blade breaks} &0.80 \\
\text{4-blade breaks}	& 0.85  \\
\hline
\end{tabular}
\label{table6}
\end{table}

\subsection{Simulations Results of Fault Diagnosis and Accommodation}
\begin{figure}[tp]	\centering
 \includegraphics[width=8.5cm,height=30cm,keepaspectratio]{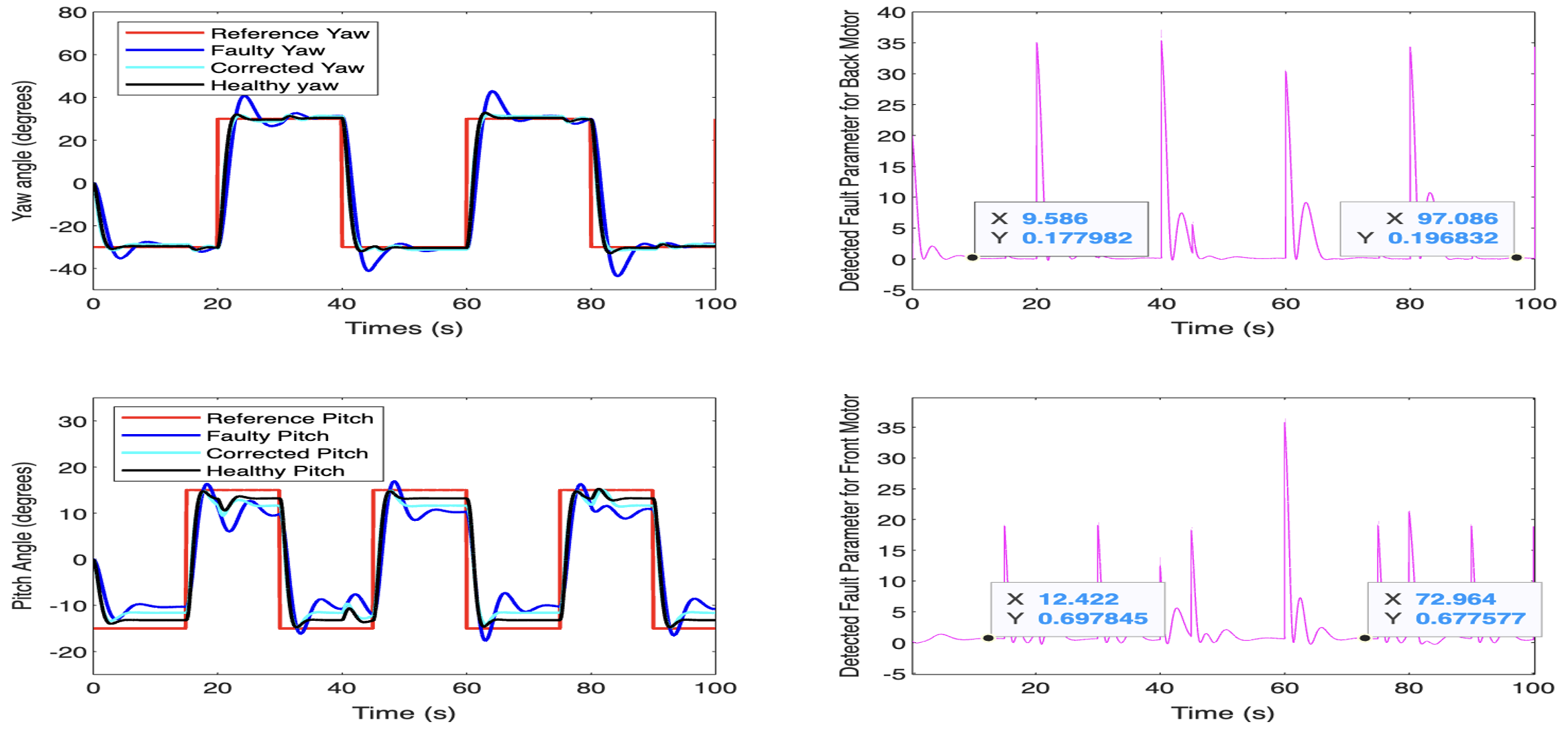}
	\caption{Fault detection and accommodation for Aero 2 2DOF model with $[\gamma_1 \quad\gamma_2]=[0.7\quad0.7]$}
	\label{image7}
\end{figure}

\cref{image7} provide the result of the fault diagnosis and accommodation technique utilized on the system model. The simulation was carried out with MATLAB/SIMULINK, and the lost of control effectiveness that were injected into the system are: $\Gamma=[0.7 \quad 0.7]^{'}$. Front and back rotors in \cref{image7} relate to Motor0 and Motor1, respectively.

	\section{Discussion}
	\label{sec4}
\subsection{Fault Analysis}
Fault analysis was carried out on the real aerospace system. \cref{image6}, \cref{table2}, and \cref{table3} provide results on the trajectory tracking performance of the 2DOF helicopter. Experiments with different fault settings indicated that while the transient response of the pitch tracking slows down, that of the yaw tracking almost remained the same. Also, the introduction of fault into the system tends to increase the SSE for the tracking of both pitch and yaw angles. Furthermore, while the overshoot in yaw tracking significantly increases with all faults, that of the pitch tracking decreases with the exception of 1 blade break on Motor0, whose pitch overshoot was noticeably higher than that of other faults.

An increase in the vibration of the system was observed as the fault worsened. To examine this closely, the SD for the steady state values of the pitch and yaw angles as well as the SD for the steady state values of Motor0 and Motor1 values were recorded. \cref{table4} and \cref{table5} provide the recorded values. The results showed that the SD of the pitch and yaw angles at steady state tends to increase with fault. Additionally, the SD value is comparatively higher when there is 1 blade break on Motor0 propeller, while the blades on Motor1 propeller were fully healthy. Moreover, the SD values for Motor0 and Motor1 steady state values are higher for 1 blade break fault. These can be attributed to the imbalance and asymmetry that arises as a result of such faults.

Open loop system response of the system further illustrated that the vibration of the system increases as the fault worsens. \cref{table6} shows that the natural frequency increases as the as the fault worsens.

Although a model-based method was used for fault diagnosis in this research, the observation that frame vibrations increase with such faults, and the possibility of using sensors like accelerometers to measure such vibrations, suggests that signal processing methods could be employed for diagnosing the system.

\subsection{Fault Diagnosis and Accommodation}
\cref{image7} illustrates the performance of the fault diagnosis and accommodation methods employed on the system model. In addition to estimating the injected fault in the system, the estimated fault parameters were able to gauge the magnitude of the coupling effects affecting the system's operation whenever there was a change of reference for one of the motions. These estimates are depicted by the spikes in the fault parameters.  Although a loss of effectiveness of $\gamma=0.7$ was injected in each rotor, the detection accuracy of the fault on the front rotor (Motor0) is relatively higher than that of the back rotor (Motor1). This could be attributed to the coupling effect that exist between the two motions. However, it could be seen that after readjustment with the estimated fault information, the control input did well in regaining the performance of the system when it was fully healthy.


	\section{Conclusions}
	\label{sec5}
In the event of faults in system components such as actuators, sensors, and the frame, conventional feedback control systems may result in poor performance or instability of the system. This problem is particularly concerning for safety-critical systems such as aircraft. Although essential for fault diagnosis and accommodation, not much has been done in analyzing the performance of a real aerospace system under different fault conditions. Therefore, this work presents the performance analysis of a 2DOF helicopter with varying fault conditions on its propeller blades. Results from the experiment show degradation in performance and an increase in vibration due to faults, especially if the faults lead to imbalance and asymmetry of the system. Although model-based fault diagnosis and accommodation methods were proposed to maintain satisfactory performance when a fault occurs, the ability to use sensors to measure fault-induced vibrations suggests that signal processing fault diagnosis methods would be effective in determining the occurrence of a fault, its location, and its severity.

 
	\label{Bibliography}
	\bibliographystyle{IEEEtran}
	\bibliography{Bibliography}
	
\end{document}